\documentclass[aps,prb,twocolumn,groupedaddress,showpacs]{revtex4-1}
\usepackage{graphics}
\usepackage{graphicx}
\usepackage{amsmath}
\usepackage{amssymb}
\usepackage{amsthm}
\usepackage{epsfig}
\usepackage{dsfont}
\begin{document}
\title{Study of the skin effect in superconducting materials}
\author{Jacob Szeftel$^1$}
\email[corresponding author :\quad]{jszeftel@lpqm.ens-cachan.fr}
\author{Nicolas Sandeau$^2$}
\author{Antoine Khater$^3$}
\affiliation{$^1$ENS Cachan, LPQM, 61 avenue du Pr\'esident Wilson, 94230 Cachan, France}
\affiliation{$^2$Aix Marseille Universit\'e, CNRS, Centrale Marseille, Institut Fresnel UMR 7249, 13397, Marseille, France}
\affiliation{$^3$Universit\'e du Maine, UMR 6087 Laboratoire PEC, F-72000 Le Mans France}
\begin{abstract}
The skin effect is analyzed  to provide the numerous measurements of the penetration depth of the electromagnetic field in superconducting materials with a theoretical basis. Both the normal and anomalous skin effects are accounted for within a single framework. The emphasis is laid on the conditions required for the penetration depth to be equal to London's length,  which enables us to validate an assumption widely used in the interpretation of all current experimental results.
\end{abstract}
\pacs{74.25.Fy,74.25.Ha}
\keywords{skin effect}
\maketitle
		\section{introduction}
	Superconductivity is characterized by two prominent properties\cite{par,ash,gen,sch,tin}, namely persistent currents in vanishing electric field and the Meissner effect\cite{mei}, which latter highlights the rapid decay of an applied magnetic field in bulk matter in a superconductor. Early insight into the Meissner effect was achieved thanks to London's equation\cite{lon}
$$B+\mu_0\lambda^2_L\textrm{curl} j=0\quad,$$
where $\mu_0,j,\lambda_L$ stand for the magnetic permeability of vacuum, the persistent current, induced by the magnetic induction $B$ and London's length, respectively. London's equation, combined with those of Newton and Maxwell, entails\cite{lon,par,ash,gen,sch,tin} that the penetration depth of the magnetic field is equal to $\lambda_L$
\begin{equation}
\lambda_L=\sqrt{\frac{m}{\mu_0\rho e^2}}\quad,
\end{equation}
where $e,\quad m,\quad\rho$ stand for the charge, effective mass and concentration of superconducting electrons. Thus the   measurement of $\lambda_L$ is all the more important, since there is no other experimental access to $\rho$.\par
	 Pippard\cite{pip1,pip,pip2} carried out the first measurements of electromagnetic energy absorption at a frequency $\omega\approx 10GHz$ in superconducting $Sn$, containing impurities, and interpreted his results within the framework of the anomalous skin effect. In normal conductors, the real part of the dielectric constant $\epsilon_R(\omega)$ being negative for $\omega<\omega_p$, where $\omega_p\approx 10^{16}Hz$ stands for the plasma frequency, causes the electromagnetic field to remain confined within a thin layer of frequency dependent thickness $\delta(\omega)$, called the skin depth, and located at the outer edge of the conductor. $\delta$ is well known\cite{jac,bor} to behave like $\omega^{-1/2}$ at low frequency and to reach, in very pure metals at low temperature, a $\omega$-independent, lower bound $\delta_a$, characteristic of the anomalous skin effect\cite{reu,cha}. Actually, in the wake of Pippard's work, all current determinations of $\lambda_L$, made in superconducting materials\cite{h3,h4,h2,h1,gor,har}, including high $T_c$ compounds ($T_c$ stands for the critical temperature), consist of measuring the penetration depth of the electromagnetic field at frequencies $\omega\in\left[10MHz,100GHz\right]$, while assuming $\lambda_L=\delta_a$.\par
 	As the latter assumption has hardly been questioned, the main purpose of this work is to ascertain its validity by working out a comprehensive analysis of the skin effect, including both the usual and anomalous cases. The treatment of the electrical conductivity at finite frequency in a superconducting material runs into further difficulty, because, according to the mainstream model\cite{ash,par,gen,tin,sch}, the conduction electrons make up, for $T< T_c$, a two-component fluid, comprising normal and superconducting electrons. This work, which is intended at deriving the respective contributions of the two kinds of electrons, is based solely on Newton and Maxwell's equations.\par
  The outline is as follows: Sections II deals with the skin effect; the results are used to work out the conduction properties of the two-fluid model in Section III, while contact is made with the experimental results in Section IV. The conclusions are given in Section V.\par
\begin{figure}
\includegraphics*[height=5 cm,width=7.5 cm]{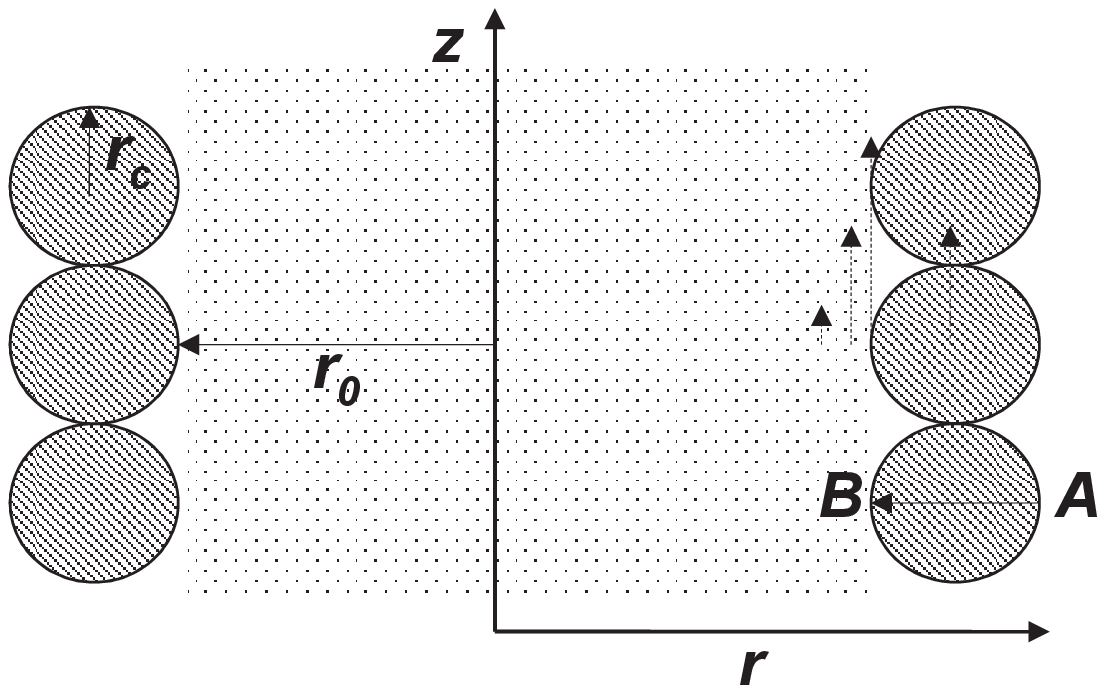}
\caption{Cross-section of the superconducting sample (dotted) and the coil (hatched); $E_\theta$ and $j_\theta$ are both normal to the unit vectors along the $r$ and $z$ coordinates; vertical arrows illustrate the $r$ dependence of $B_z(r)$; $r_c$ has been magnified for the reader's convenience; Eq.(\ref{coil}) has been integrated from $A$ ($B_z(r_0+2r_c)=0$) to $B$}\label{Bzr}
\end{figure}
	\section{skin effect}
	Consider as in Fig.1 a superconducting material of cylindrical shape, characterized by its symmetry axis $z$ and radius $r_0$ in a cylindrical frame with coordinates ($r,\theta,z$). The material is taken to contain conduction electrons of charge $e$, effective mass $m$, and total concentration $\rho$. It is subjected to an oscillating electric field $E(t,r)=E_\theta(r)e^{i\omega t}$, with $t$ referring to time. As $E_\theta(r)$ is normal to the unit vectors along the $r$ and $z$ coordinates, there is $\textrm{div} E=0$.
$E$ induces a current $j(t,r)=j_\theta(r)e^{i\omega t}$ along the field direction, as given by Newton's law
\begin{equation}
\label{newt}
\frac{dj}{dt}=\frac{\rho e^2}{m}E-\frac{j}{\tau}\quad,
\end{equation}
where $\frac{\rho e^2}{m}E$ and $-\frac{j}{\tau}$ are respectively proportional to the driving force accelerating the conduction electrons and a generalized friction term, which is non zero in any superconducting material provided $\frac{dj}{dt}\neq0$.\par
	 The existence of a \textit{friction force} in superconductors, carrying an \textit{ac current}, is known experimentally (see\cite{sch} p.4, $2^{nd}$ paragraph, line $9$). For example, the measured ac conductivity for the superconducting phase of $BaFe_2(As_{1-x}P_x)_2$ has been found (see\cite{h1} p.1555, $3^{rd}$ column, $2^{nd}$ paragraph, line $11$) to be $\approx .03\sigma_n$, where $\sigma_n$ stands for the normal conductivity\cite{foo} measured just above the critical temperature $T_c$. However because the current is carried by electrons, making up either a BCS state\cite{bar} or a Fermi gas\cite{ash} in a superconducting and normal metal, respectively, the physical sense of $\tau$ in Eq.(\ref{newt}) for superconductors may be different from that given by the Drude model\cite{ash} for a normal metal. To understand this difference and to model the new $\tau$, we shall next work out the equivalent of Ohm's law for a superconducting material when submitted to an electric field.\par
	The superconducting state, carrying no current in the absence of external fields, is assumed to comprise two subsets of equal concentration $\rho/2$, moving in opposite directions with respective mass center velocity $v,-v$, which ensures $j=0,\quad p=0$, where $p$ refers to the average electron momentum. Under a driving field $E$, an ensemble $\delta\rho/2$ of electrons is transferred from one subset to the other, so as to give rise to a finite current $j=\delta\rho ev=e\delta p/m$, where $\delta p$ stands for the electron momentum variation. The generalized friction force is responsible for the reverse mechanism, whereby electrons are transferred from the majority subset of concentration $\frac{\rho+\delta\rho}{2}$ back to the minority one ($\frac{\rho-\delta\rho}{2}$). It follows from flux quantization and the Josephson's effect\cite{par,ash,gen,sch,tin,jos} that the elementary transfer process involves a pair\cite{coo} rather than a single electron.
Hence if $\tau^{-1}$ is defined as the transfer probability per unit time of one electron pair, the net electron transfer rate is equal to $\frac{\rho+\delta\rho-(\rho-\delta\rho)}{2\tau}=\frac{\delta\rho}{\tau}$.
By virtue of Newton's law, the resulting generalized friction term is $mv\delta\rho/\tau=\delta p/\tau\propto j/\tau$, which validates Eq.(\ref{newt}). Furthermore for $\omega\tau<<1$, the inertial term $\propto\frac{dj}{dt}$ in Eq.(\ref{newt}) is negligible, so that we can write the equivalent of Ohm's law for the superconducting material as
\begin{equation}
j=\sigma E\quad,\quad\sigma=\frac{\rho e^2\tau}{m}\quad.
\end{equation}
Thus both Ohm's law and $\sigma$ are seen to display\cite{ash} the same form in normal and superconducting metals, as well.\par
$E$ induces a magnetic induction $B(r,t)=B_z(r)e^{i\omega t}$, parallel to the $z$ axis. $B$ is given by the Faraday-Maxwell equation as
\begin{equation}
\label{Bz}
-\frac{\partial B}{\partial t}=\textrm{curl}E=\frac{E}{r}+\frac{\partial E}{\partial r}\quad.
\end{equation}
The displacement vector $D$, is parallel to $E$ and is defined as
\begin{equation}
D=\epsilon_0 E+\rho e u\quad,
\end{equation}
where $\epsilon_0,\quad u$ refer to the electric permittivity of vacuum and displacement coordinate of the conduction electron center of mass, parallel to $E$. The term $\rho e u$ represents the polarization of conduction electrons\cite{fo2}. Because $\textrm{div} E=0$ entails that $\textrm{div} D=0$, Poisson's law warrants the lack of charge fluctuation around $\rho e$. Thence since there is by definition $j=\rho e\frac{du}{dt}$, the displacement current reads
\begin{equation}
\frac{\partial D}{\partial t}=j+\epsilon_0\frac{\partial E}{\partial t}\quad.
\end{equation}
Finally the magnetic field $H(t,r)=H_z(r)e^{i\omega t}$, parallel to the $z$ axis, is given by the Amp\`ere-Maxwell equation as
\begin{equation}
\label{Hz}
\textrm{curl}H=-\frac{\partial H}{\partial r}=j+\frac{\partial D}{\partial t}=2j+\epsilon_0\frac{\partial E}{\partial t} \quad.
\end{equation}
 Replacing $E(t,r),j(t,r),B(t,r),H(t,r)$ in Eqs.(\ref{newt},\ref{Bz},\ref{Hz}) by their time-Fourier transforms $E_\theta(\omega,r),j_\theta(\omega,r),B_z(\omega,r),H_z(\omega,r)$, while taking into account
$$B_z\left(\omega,r\right)=\mu\left(\omega\right)H_z\left(\omega,r\right)\quad,$$
where $\mu\left(\omega\right)=\mu_0\left(1+\chi_s\left(\omega\right)\right)$ ($\chi_s\left(\omega\right)$ is the magnetic susceptibility of superconducting electrons at frequency $\omega$) yields 
\begin{equation}\label{fou2}
\begin{array}{l}
E_\theta\left(\omega,r\right)=\frac{1+i\omega\tau}{\sigma}j_\theta\left(\omega,r\right)\\
i\omega B_z\left(\omega,r\right)=-\left(\frac{E_\theta\left(\omega,r\right)}{r}+\frac{\partial E_\theta\left(\omega,r\right)}{\partial r}\right)\\
\frac{\partial B_z\left(\omega,r\right)}{\partial r}=-\mu\left(\omega\right)\left(2j_\theta\left(\omega,r\right)+i\omega\epsilon_0E_\theta\left(\omega,r\right)\right)
\end{array}
\end{equation}
Eliminating $E_\theta\left(\omega,r\right),j_\theta\left(\omega,r\right)$ from Eqs.(\ref{fou2}) gives
\begin{equation}
	\label{skin}
\frac{\partial^2 B_z\left(\omega,r\right)}{\partial r^2}=\frac{B_z\left(\omega,r\right)}{\delta^2(\omega)}-\frac{\partial B_z\left(\omega,r\right)}{r\partial r}\quad.
\end{equation}
The skin depth $\delta$ and plasma frequency $\omega_p$ are defined\cite{ash,jac,bor} as
$$\begin{array}{l}	\delta(\omega)=\frac{\lambda_L}{\sqrt{\left(1+\chi_s\left(\omega\right)\right)\left(\frac{2i\omega\tau}{1+i\omega\tau}-\frac{\omega^2}{\omega^2_p}\right)}}\quad,\\
\omega_p=\sqrt{\frac{\rho e^2}{\epsilon_0m}}
\end{array}$$
  Note that the above formula of $\delta(\omega)$ retrieves indeed both, the usual\cite{jac,bor} expression $|\delta|=\frac{1}{\sqrt{2\mu_0\sigma\omega}}$, valid for $\omega\tau<<1$, and the $\omega\tau>>1$ limit $\delta_a=\frac{\lambda_L}{\sqrt{2}}$, typical of the anomalous skin effect\cite{reu,cha} and widely used in the interpretation of the experimental work\cite{pip1,pip,pip2,h3,h4,h2,h1,gor,har}.\par
\begin{figure}
\includegraphics*[height=6 cm,width=6 cm]{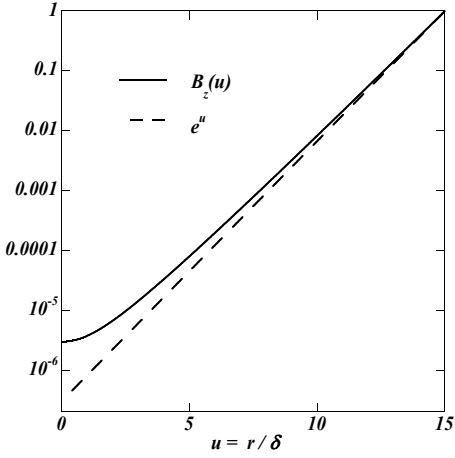}
\caption{Semi-logarithmic plots of $B_z(u),e^u$.}\label{f2}
\end{figure}
   As Eqs.(\ref{fou2}) make up a system of $3$ linear equations in terms of $3$ unknowns $j_\theta,E_\theta,B_z$, there is a single solution, embodied by Eq.(\ref{skin}). The solution of Eq.(\ref{skin}), which has been integrated over $r\in \left[0,r_0\right]$ with the initial condition $\dfrac{dB_z}{dr}\left(r=0\right)=0$, is a Bessel function, having the property $B_z(r)\approx e^{r/\delta(\omega)}$ if $r>>|\delta(\omega)|$, as illustrated in Fig.\ref{f2}.\par
	\section{the two-fluid model}
	The total current $j_t$ reads as 
		$$j_t=j_n+j_s=(\sigma_n+\sigma_s)E\quad,$$
where $j_n,\sigma_n=\frac{\rho_n e^2\tau_n}{m_n}$ ($j_s,\sigma_s=\frac{\rho_s e^2\tau_s}{m_s}$) designate the normal (superconducting) current and conductivity, with $\rho_n,\tau_n,m_n$ ($\rho_s,\tau_s,m_s$) being the concentration, decay time of the kinetic energy, associated with $j_n$ ($j_s$), and effective mass of normal (superconducting) electrons. Replacing $j_\theta$ in Eqs.(\ref{fou2}) by the expression $j_t$ hereabove leads to the following expression of the skin depth for the practical case $\omega<<\omega_p$ and $\chi_s<<1$
\begin{equation}
\label{skin3}
\delta^{-2}=2i\mu_0\omega\left(\frac{\sigma_n}{1+i\omega\tau_n}+\frac{\sigma_s}{1+i\omega\tau_s}\right)\quad.
\end{equation}
The inequalities $\rho_n<<\rho_s,\tau_n<<\tau_s,\sigma_n<<\sigma_s$ can be inferred from observation\cite{h3,h4,h2,h1,gor,har} to hold at $T$ well below $T_c$, so that $\delta$ in Eq.(\ref{skin3}) is recast finally as
$$\begin{array}{l}
|\delta\left(\omega<<\tau_s^{-1}\right)|=\frac{1}{\sqrt{2\mu_0\sigma_s\omega}}=\frac{\lambda_L}{\sqrt{2\omega\tau}}\quad,\\
\delta\left(\omega>>\tau_s^{-1}\right)=\sqrt{\frac{m_s}{2\mu_0\rho_s e^2}}=\frac{\lambda_L}{\sqrt{2}}\quad,
\end{array}$$
where both limits $\omega\tau_s<<1,\quad \omega\tau_s>>1$ are seen to correspond to the usual and anomalous skin effect, respectively. Besides it is concluded that the conduction properties, at $\omega\neq 0$ and $T$ well below $T_c$, are assessed solely by the superconducting electrons, while the normal ones play hardly any role.
	\section{comparison with experiment}
	The experiments, performed at $\omega\approx 10GHz$, give access to the imaginary part of the complex impedance of the resonant cavity, containing the superconducting sample, equal to $\mu_0\omega l_p$, where $l_p$ designates the penetration depth of the electromagnetic field, i.e. $l_p=\delta$. Moreover at $\omega\approx 10MHz$, the cavity is to be replaced by a resonant circuit, combining a capacitor and a cylindrical coil of radius $r_0$. The sample is inserted into the coil, which is flown through by an oscillating current $I_0(\omega)e^{i\omega t}$. The coil is made up of a wire of length $l$ and radius $r_c$ (see Fig.\ref{Bzr}). Since the relationship between the penetration depth $l_p$ and the observed self inductance of the coil $L$ is seemingly lacking, we first set out to work out an expression for $L$.\par
	 Applying Ohm's law yields
\begin{equation}
\label{ohm}\begin{array}{l}
-l\left(E_a(\omega)+E_\theta(\omega,r_0)\right)=RI_0(\omega)\Rightarrow\\
E_\theta(\omega,r_0)=\frac{U(\omega)-RI_0(\omega)}{l}
\end{array}\quad,
\end{equation}
where $E_a(\omega)e^{i\omega t},E_\theta(\omega,r)e^{i\omega t},Ue^{i\omega t}=-lE_a(\omega)e^{i\omega t},R$ are the applied and induced electric fields, both normal to the $r,z$ axes, the voltage drop throughout the coil and its resistance, respectively $\left(E_a(\omega),E_\theta(\omega,r),U(\omega)\in\mathds{C}\right)$. Besides $E_\theta(\omega,r_0)$ is obtained from Eq.(\ref{fou2}) as
\begin{equation}
\label{eth}E_\theta(\omega,r_0)=-i\omega\delta(\omega)B_z(\omega,r_0)\quad.
\end{equation}
where $E_\theta(\omega,r\rightarrow r_0)\approx E_\theta(\omega,r_0)e^{\frac{r-r_0}{\delta(\omega)}}$. Working out $B_z(\omega,r_0)$ in Eq.(\ref{eth}) requires to solve the Amp\`ere-Maxwell equation for $B_z(\omega,r)e^{i\omega t}$ inside a cross-section of the coil wire
\begin{equation}
	\label{coil}
\frac{\partial B_z}{\partial r}=-\mu_0\left(2j_c+i\epsilon_0\omega E_c\right) \quad,	
\end{equation}
where $j_c(\omega)=\frac{I_0(\omega)}{\pi r_c^2}$ and $E_c(\omega)=E_a(\omega)+E_\theta(\omega,r_0)$ are both assumed to be $r$-independent. Moreover integrating Eq.(\ref{coil}) for $r\in\left[r_0+2r_c,r_0\right]$ with the boundary condition $B_z(\omega,r_0+2r_c)=0,\forall t$ (see Fig.\ref{Bzr}), while taking advantage of Eq.(\ref{ohm}), yields
\begin{equation}
\label{Bzt}
B_z(\omega,r_0)=2\mu_0\left(\frac{2}{\pi r_c}-\frac{i\epsilon_0\omega r_cR}{l}\right)I_0(\omega)\quad.
\end{equation}
Finally it ensues from the definition of $L$ and Eq.(\ref{Bzt}) that
$$LI_0=2\pi Re\left(\int_0^{r_0}B_z(\omega,r_0)e^{\frac{r-r_0}{\delta}}rdr\right)\Rightarrow L\approx 2^{\frac{5}{2}}\mu_0|\delta|\frac{r_0}{r_c},$$
where $Re$ means real part and $\delta=\lambda_L/\sqrt{2i\omega\tau}$ for $\omega\tau<<1$.\par
 Both imaginary parts of the coil and cavity impedance are thus found to be equal to $C(\omega)\omega|\delta(\omega)|$, where $C(\omega)$ is an unknown coefficient depending on the experimental conditions. As $C(\omega)$ is likely to be $T$ independent at least inside the limited range $0<T<T_c$, this hurdle can be overcome by plotting the dimensionless ratio $\left|\frac{\delta(T)}{\delta(T_c)}\right|$ versus $T\in\left[0,T_c\right]$, where $\delta(T)$ is the skin depth measured at two frequencies far apart from each other, e.g. $\omega_1=10GHz,\quad\omega_2=10MHz$. The reason for suggesting such an experimental procedure is that the value of $\delta(T_c)$ is well known, because $\delta(T_c)=\frac{1}{\sqrt{2\mu_0\sigma_n(T_c)\omega}}$ and the normal conductivity $\sigma_n(T_c)$ can be measured with great accuracy. Consequently two cases should be considered :\\
\textit{i)} the two curves representing $\left|\frac{\delta(T\in\left[0,T_c\right])}{\delta(T_c)}\right|$, corresponding to $\omega_1,\quad\omega_2$ cannot be distinguished from each other, which entails that $\delta(T<T_c)\propto1/\sqrt{\omega_{i=1,2}}$ even for $T\rightarrow T_c^-$. As this conclusion implies also that $\omega_{i=1,2}\tau_s<<1$, this case is likely to be observed in high-$T_c$ superconducting alloys, wherein $\tau_s$ is bound to be relatively short because of numerous impurities;\\
\textit{ii)} conversely the two curves turn out to be conspicuously different, which points toward the limit $\omega_1\tau_s>>1,\quad\omega_2\tau_s<<1$, characterized by $\delta(\omega_1,T<T_c)=\frac{\lambda_L(T)}{\sqrt{2}},\quad \delta(\omega_2,T<T_c)=\frac{\lambda_L(T)}{\sqrt{2\omega_2\tau_s}}$. Since a very long $\tau_s$ is required, the good candidates are likely to be chosen among the elementary superconducting metals of very high purity\cite{pip1,pip,pip2}. However, because of $\rho_s(T\rightarrow T_c^-)\rightarrow0$, the $\omega$ dependence of $\delta(T<T_c)$ will switch from $\delta(T)$ independent from $\omega$ for $T$ well below $T_c$ to $\delta\propto\omega^{-1/2}$ for $T\rightarrow T_c^-$.\par 
	Because $\tau_s$ is not known, $\lambda_L(T)$ cannot be measured in the limit $\omega\tau_s<<1$. Therefore it is suggested to work at higher frequencies, such that $\omega>>1/\tau_s,\omega<<\omega_p$, because $\delta(\omega)=\lambda_L/\sqrt{2}$ is independent from $\tau_s$ in that range. Typical values $\tau_s\approx 10^{-11}s,\omega_p\approx 10^{16}Hz$ would imply to measure light absorption in the IR range. Then for an incoming beam being shone at normal incidence on a superconductor of refractive index $\tilde{n}\in\mathds{C}$, the absorption and reflection coefficients $A,R$ read\cite{jac,bor}
 $$ A=1-R=1-\left|\frac{1-\tilde{n}}{1+\tilde{n}}\right|^2\quad.$$
 The refractive index $\tilde{n}$ and the complex dielectric constant $\epsilon=\epsilon_R+i\epsilon_I$, conveying the contribution of conduction electrons, are related\cite{jac,bor} by
 $$\tilde{n}^2=\frac{\epsilon}{\epsilon_0}=1-\frac{\left(\omega_p/\omega\right)^2}{1-i/(\omega\tau_s)}\quad.$$
At last we get
$$\lambda_L=\frac{2}{\mu_0c\sigma_s A}\quad,\quad\tau_s=\mu_0\sigma_s\lambda_L^2\quad,$$
where $c$ refers to light velocity in vacuum. 
	\section{conclusion}
	The skin effect has been analyzed in a single framework, to account for both the cases of the usual and anomalous skin effect. The calculation of the skin depth has then been applied to the study of electromagnetic wave propagation in superconducting materials and to the interpretation of skin depth measurements. An experiment has been proposed to measure London's length in dirty superconductors.


\begin{thebibliography}{00}
\bibitem{fo2}
the contribution of electrons bound to the nucleus can be taken into account, by replacing hereafter $\epsilon_0$ by $\epsilon_0+\alpha$, where $\alpha$ refers to their polarisability
\end{thebibliography}
\end{document}